\documentclass[12pt,preprint]{aastex}
\slugcomment{{\sc Accepted to ApJ:} January 15, 2013} 
\usepackage{natbib}

\shorttitle{Linear Polarization in Classical Be Stars}
\shortauthors{Halonen \& Jones}

\begin{document}

\title{On the Intrinsic Continuum Linear Polarization of Classical Be Stars during Disk Growth and Dissipation}		
\author{Robbie J. Halonen \& Carol E. Jones}
\affil{Department of Physics and Astronomy, The University of Western Ontario, London, ON, Canada, N6A 3K7}
\email{rhalonen@uwo.ca}

\begin{abstract}
We investigate the intrinsic continuum linear polarization from axisymmetric density distributions of gas surrounding classical Be stars during the formation and dissipation of their circumstellar disks. We implement a Monte Carlo calculation of the Stokes parameters with the use of the non-LTE radiative transfer code of \citet{sig07} to reproduce the continuous polarimetric spectra of classical Be stars. The scattering of light in the nonspherical circumstellar envelopes of classical Be stars produces a distinct polarization signature that can be used to study the physical nature of the scattering environment. In this paper, we highlight the utility of polarimetric measurements as important diagnostics in the modeling of these systems. We illustrate the effects of using self-consistent calculation of the thermal structure of the circumstellar gas on the characteristic wavelength-dependence of the polarization spectrum. In showing that the principal features of the polarization spectrum originate from different parts of the disk, we emphasize the capability of polarimetric observations to trace the evolution of the disk on critical scales. We produce models that approximate the disk formation and dissipation periods and illustrate how the polarimetric properties of these systems can have a pivotal role in determining the mechanism for mass decretion from the central star.
\end{abstract}


\section{Introduction}

In his treatise on the transfer of radiation in stellar atmospheres, \citet{cha46} predicted the polarization of continuous light in the emergent radiation from early-type stars. He theorized that the polarization from light scattered in the darkened limbs of stellar atmospheres could be detectable under favourable conditions. While the early observational polarimetry is widely recognized for the serendipitous discovery of polarization from scattering by interstellar dust, it also confirmed that some stars emit intrinsically polarized light. These observations suggested a different source for the detected polarization then that of the hypothesis that spurred the investigation; instead, the polarigenic mechanism was determined to be the scattering of radiation in the circumstellar material enveloping the stars \citep[see, for example,][]{coy69,zel72}. Since then, the observation of stellar phenomena through polarimetric observation has yielded important findings for many different types of stars. For an overview of the history and application of polarimetry in stellar astronomy, we refer interested reader to the excellent textbook by \citet{cla10}.

It is now known that the intrinsic polarization from scattering is an important property for investigating the nature of celestial objects. Particularly, the polarization signature of the light from an observed target contains information regarding the geometric properties of the system. Clues to the astronomical geometry of sources of scattered light may be ascertained by examining the orientation associated with the observed polarized light. In the case of stars with circumstellar environments, polarimetric observations of stars provide a means of examining the surrounding material without having to resolve the object. Because the polarigenic mechanism responsible for the detected polarized light is intrinsic to the surrounding envelope, the signature is a separate observational feature that characterizes the scattering material. As such, stellar polarimetry provides a direct means for probing the scattering environments around particular stars.

In this paper, we examine the characteristics of the linearly polarized continuous light produced in classical Be stars. Before proceeding, it is necessary to distinguish between objects that are designated as \textit{classical} and those that are not. The general term \textit{Be star} describes any B-type star that exhibits spectral line emission. This description commonly includes, but is not limited to, classical Be stars, Herbig Be stars, B[e] stars and luminous blue variables. The term \textit{classical Be star} refers to an object for which there is observational evidence that it possesses, or has possessed at some point in time, a circumstellar disk of gas orbiting the star in Keplerian fashion \citep{por03}.

The prototypical classical Be star is characterized by observational features that reflect the interaction between the radiation emitted by a hot massive star and the cooler material present in its circumstellar environment. It is now accepted that this material exists predominantly in the form of a thin, equatorial, decretion disk of ionized gas with minimal radial outflow. This is confirmed by interferometric observations which resolve the disk of nearby classical Be stars \citep{qui97} and spectroastrometric observations which verify that the disk rotation is Keplerian \citep{whe12}. In addition to a linear polarization signature, the distinct set of features observed in classical Be stars include a prominent emission-line spectrum, an excess of continuum emission and periodic line profile variations. Due to changing physical conditions in the disk, the relative intensity of these observational properties may vary on timescales that range from days to decades. These observed features may even disappear, a phenomenon that is interpreted as the complete dissipation of the surrounding gas, and reappear after a period when the disk reforms.

The physical description of the Keplerian disk adheres well to the steady-state viscous disk model \citep{lee91,oka01} which can explain numerous features of classical Be stars \citep{car11}. Despite extensive observation and study, however, the processes involved in the formation of the viscous disks remain unidentified. While several mechanisms have been proposed \citep[see][]{owo06}, none rigorously satisfy all of the observational requirements. Rapid rotation is presumed to be vital to the process as classical Be stars rotate more rapidly than average B-type stars. Typical rotational velocities are measured at about 70 to 80 \% of the critical velocities \citep{por96}, although this number may be systematically underestimated due to the inherent nature of these systems \citep{tow04}. Mass ejection from the central star may be driven by non-radial pulsations \citep{riv03,cra09} or binary companions \citep{har02}. However, a viable process for providing sufficient angular momentum to achieve a tenable Keplerian disk remains elusive. The collaborative efforts of observers and theorists are crucial to resolving this long-standing challenge of classical Be stars.

The motivation of theoretical studies is to identify the processes involved in the evolution of stellar decretion disks and to improve our understanding of the interaction between massive stars and their circumstellar environments. In order to do so, it is important to accurately determine the physical properties of these disks through observation and modeling. With particular regards to this study, non-zero polarization levels constitute a key observable for investigating the nature of classical Be stars, particularly for probing the geometric properties of the circumstellar environment. With modern instrumentation, polarimetric measurements with high spectral and temporal resolution are achievable. It is critical that realistic and well-understood theoretical models are available to interpret these observations and to underlie their relevance.

In this paper, we present the implementation of a Monte Carlo calculation of the Stokes parameters using the self-consistent solution of the non-LTE radiative transfer code of \citet{sig07} to investigate the physical properties of the circumstellar disks of classical Be stars. This study provides much-needed theoretical consideration of classical Be stars for an important astronomical technique and reinforces the merits of polarimetry in tracing the evolution of circumstellar disks. We aim to improve the visibility of polarimetry within the community by highlighting the utility of this observing method in investigating circumstellar environments and by emphasizing an understanding of the underlying physical properties which yield the observed polarization signature.  We also use models representing the growth and dissipation of a disk surrounding classical Be stars to demonstrate how polarimetric observables are ideally suited for tracing physical changes in the circumstellar disk. These polarimetric diagnostics will be important in understanding mass decretion in classical Be stars and divining the origin of circumstellar disks. 

We have organized the paper as follows: in Section 2, we review the basics of polarization and the Stokes parameters; in Section 3, we describe the computational procedures relevant to this investigation; in Section 4, we discuss an analysis of varying the parameters used to evaluate the polarization levels from scattering in the circumstellar gas; in Section 5, we examine the changes to the polarization spectrum while simulating the processes of disk growth and dissipation; and lastly, in Section 6, we summarize our results and discuss our ongoing research.

\section{Polarization and the Stokes Parameters}

Thomson scattering is the primary source of intrinsic continuum polarization in classical Be stars. When unpolarized light undergoes scattering by free electrons, it becomes linearly polarized perpendicular to the plane containing the incident and scattered radiation. When Thomson scattering occurs in a source that appears spherically symmetric on the plane of the sky, the distribution of polarizing planes is uniform, resulting in the complete cancellation of vibrations from orthogonal directions. In a classical Be star, the projection of the circumstellar disk is spherically symmetric when the rotational axis of the star coincides with the observing line-of-sight. In other words, classical Be stars should exhibit no net polarization when viewed pole-on. When the rotational axis of a classical Be star is inclined with respect to the direction of an observer, the projection of the disk on the observation plane becomes non-radially symmetric and the complete cancellation of the vibrations does not occur. Thus, classical Be stars exhibit net polarization when viewed at higher inclinations, with the polarimetric position angle cast perpendicularly to the plane of the disk. 

Among the earliest indications that the light emitted by some classical Be stars is partially polarized were the variable polarimetric measurements in $\zeta$ Tau by \citet{hal50} and $\gamma$ Cas by \citet{beh59}. Since then, observations have shown that a significant number of classical Be stars exhibit some intrinsic linear polarization of their continuous light. The degree of this polarization signature is often variable, but it has been measured to as high as a few percent of the total emitted radiation. Analysis of the intrinsic polarization distribution for a sample of 495 objects by \citet{yud00} concluded that 95\% classical Be stars exhibit polarization on a level of 0\%  $<$ p $<$ 1.5\%.  

Analyzing the behaviour of the polarized light from scattering in the ionized disk can yield valuable information regarding the nature of the medium. The importance of polarimetric observations in monitoring these systems is exhibited in several ways: the orientation of the polarization position angle provides information regarding the astronomical geometry of the disk; the relative intensity of the polarization reflects the number of scatterers in the disk; the wavelength-dependence of the polarization spectrum reveals the composition and density of the disk material; and the time-dependence of the position angle and the relative intensity exposes evolutionary trends in the physical structure of the disk. 

Although electron scattering is a wavelength-independent process, there are two effects which contribute to the wavelength-dependence observed in spectropolarimetric measurements of classical Be stars. The first effect is the dilution of the polarized light from the emission of unpolarized light by material outside of the scattering region. The addition of unpolarized light at particular wavelengths results in a decrease in the ratio of the polarized intensity to the total intensity. This decrease in the polarization level is noticeable in strong emission lines that arise largely in extended regions of the circumstellar disk, outside of the scattering region. The second effect is the attenuation of scattered light due to disk opacity. In particular, the hydrogen bound-free absorptive opacity produces a distinct sawtooth shape that is commonly observed in sufficiently strong polarization spectra of classical Be stars \citep[see, for example,][]{poe79,woo97}. 

Our consideration of the polarization from electron scattering in circumstellar environments builds on many previous theoretical investigations. Methods for predicting the continuum polarization from particular geometries have improved on the early analytic solutions for plane-parallel atmospheres derived by \citet{cha60} and \citet{col70} and the subsequent single-scattering approximations used by \citet{bro77} and \citet{poe78}, for example. Using Monte Carlo simulations, \citet{woo96a} showed that multiple scattering in axisymmetric circumstellar distributions of gas yields polarization levels that are higher than those predicted by single-scattering models. After including absorptive opacity in their model, \citet{woo96b} highlighted the importance of multiple-scattering for accurately modeling features in the continuous polarization spectrum. At present, sophisticated radiative transfer models \citep{car06b,sig07} have provided significant insight into the electron density and temperature structure of the circumstellar disk and have enabled greater understanding of the physical conditions that generate the observed characteristics of classical Be stars. 

\section{Computational Method}

\subsection{Radiative Transfer Code {\sc bedisk}}

In a series of papers, \citet{mil98,mil99a,mil99b} demonstrated the importance of using a realistic non-isothermal temperature structure in considering the predicted observables of classical Be stars. For the purposes of this work, the underlying theoretical models of the classical Be stars and their gaseous envelopes are computed using the non-LTE radiative transfer code {\sc bedisk} developed by \citet{sig07}. This computational code solves the coupled problems of statistical equilibrium and radiative equilibrium to provide a self-consistent calculation of the thermal structure of the circumstellar disk. \citet{tyc08} and \citet{jon09} illustrate the successful use of the {\sc bedisk} code in the modeling of interferometric observations of classical Be stars. 

In the computations performed by {\sc bedisk}, the energy input into the disk is assumed to originate entirely from radiation from the central star. The code employs an {\sc atlas9} LTE stellar atmosphere \citep{kur93} to calculate the stellar photoionizing radiation field in the disk. The modelled disk is assumed to be axisymmetric about the stellar rotation axis, and symmetric about the equatorial midplane. The density distribution of gas in the envelope at coordinates $(R, Z)$, where $R$ is the radial distance from the stellar rotation axis and $Z$ is the distance above the midplane, is specified by 
\begin{equation} \label{eq:dens}
 \rho(R,Z) = \rho_{0}(R)^{-n}e^{-(\frac{Z}{H})^2}.
\end{equation} 
Here, $\rho_0$ is the density at the base of the disk or where the disk encounters the stellar surface, $n$ is a power-law index and $H$ is a vertical scale height which depends on an initial value of the disk temperature. At each $R$, the gas is taken to be in vertical hydrostatic equilibrium perpendicular to the plane of the disk. At each computational grid location in the disk, the code computes the level populations for a specified set of atoms and ions and balances the heating (photoionization and collisional excitation) and cooling (recombination and collisional de-excitation) rates. 

\subsection{Multiple Scattering Code}

We employ a Monte Carlo procedure for calculating the fractional polarization of the emergent intensity due to electron scattering in the disk. The thermal solution computed by {\sc bedisk} provides the circumstellar gas model used by our multiple scattering routine. Our computations follow a straightforward algorithm for simulating the propagation of photons through the extended envelope of material that surrounds a star, similar to the procedures found in \citet{luc99} and \citet{car06a}. A complete description of our procedure is included in \citet{hal12}.

We compared the output of our code to previous analytic and numerical results as a method for ascertaining the validity of our technique. In order to verify the accuracy of our calculations of the Stokes parameters, we simulated the reflection and transmission of light by a Rayleigh scattering atmosphere. Analytic solutions of the radiative transfer equations for the intensity and polarization of the radiation emerging from such atmosphere were introduced by \citet{cha60}. Full tables of updated solutions have been published, most recently by \citet{nat09}. We simulated this problem to calculate the quantities describing the emergent radiation using our Monte Carlo scattering procedure. We ran this simulation for plane-parallel slabs bounded by a reflecting surface of varying albedoes and optical depths and compared the results for different incident and viewing angles. The comparisons showed agreement within the estimated errors between our calculations and the analytic solutions.

We also tested our code against a single-scattering plus attenuation procedure for calculating the continuum linear polarization level from a classical Be star. Both procedures employ the level populations calculated by the non-LTE radiative transfer code {\sc bedisk}. The single-scattering calculations represent an update to the work of \citep{poe78} who modeled observations of the classical Be star $\gamma$ Cas with reasonable success despite the computational limitations at the time. Our comparison showed a good agreement between the two procedures. An interesting consideration that arises from this comparison is the effect of multiple scattering in the circumstellar medium. An analysis of single-scattering versus multiple-scattering in circumstellar disks and further details of the code comparisons mentioned above are presented in \citet{hal12}.

\section{Parameter Study}

The primary focus of this work is to show the intrinsic continuum polarization levels that arise from circumstellar distributions of gas in particular axisymmetric geometries. As we described in the previous section, the radiative transfer code {\sc bedisk} solves the statistical and radiative equilibrium problems to compute a self-consistent solution for the thermal structure of a circumstellar disk constructed using equation \ref{eq:dens}. This density equation contains two free parameters which determine the initial allocation of gas surrounding the star: the density of the disk at the stellar surface $\rho_0$ and the power-law index $n$. While the assumption of a power-law density distribution limits our consideration of smaller scale disk structure, it allows for the effect of disk density as a function of radial distance to be probed with just two free parameters and represents an ideal model for studying disk growth and dissipation, as we will discuss in Section 5.  In this section, we examine the effects of varying the density on the resulting polarization spectra, and we consider the importance of the inclination of the system and the temperature structure of the disk. While investigations of these effects are not novel, they provide the necessary framework for understanding the physical changes that shape the resultant polarization spectrum. Recognizing these effects is crucial for evaluating the use of polarimetry as a diagnostic for disk evolution.

\subsection{Density Distribution}

The simplest changes to the gas density distribution in the circumstellar disk are through adjustments to the density parameters $\rho_0$. We have computed theoretical polarization spectra for a large number of disk models. In each set of models of a disk surrounding a star of a given spectral type, we vary $\rho_0$ from $1.0 \times 10^{-10}$ g cm$^{-3}$ to $1.0 \times 10^{-13}$ g cm$^{-3}$ for a power-law index of $n = 3.5$. We present the results from the set of models using the stellar properties of a representative B2V star (see Table \ref{tab:param1}) as the central star in the system under consideration. The sets of models for other spectral types exhibit qualitatively similar results. We also examined models for varying power-law indices and found the results to support the implications made from the varying base density models.

\begin{figure}
\epsscale{1.0}
\plotone{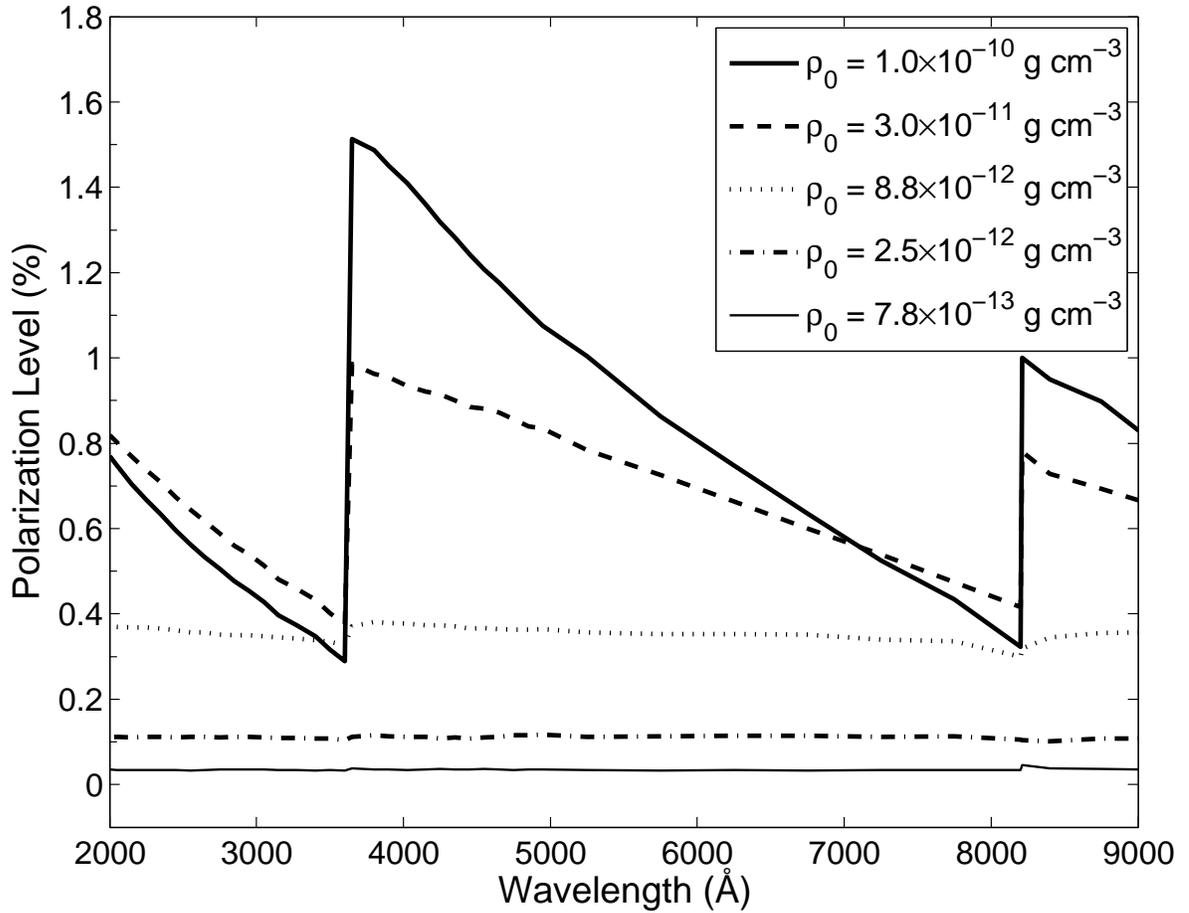}
\caption{Polarization spectra for models with circumstellar disks with varying $\rho_0$ and with $n = 3.5$. The modeled star is a B2V star viewed at an inclination of $i$ = $75^{\circ}$.}
\label{fig:pol_rho1}
\vspace{0.1in}
\end{figure}

\begin{figure}
\epsscale{1.0}
\plotone{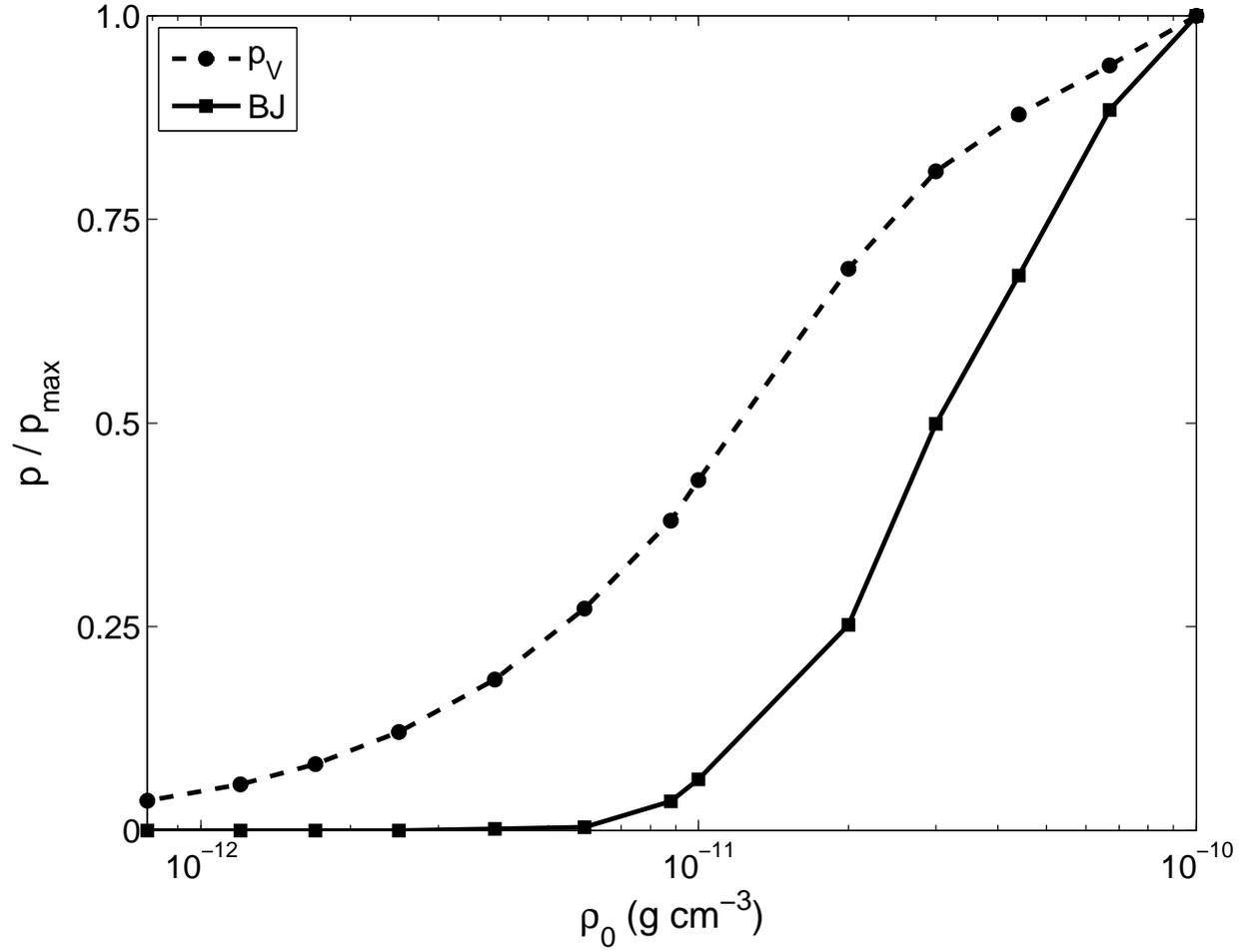}
\caption{Comparison of the growth of polarimetric features (polarization level in V-band $p_V$ and height of Balmer jump $BJ$) for models with circumstellar disks with varying $\rho_0$ and with $n = 3.5$. The modeled star is a B2V star viewed at an inclination of $i$ = $75^{\circ}$. }
\label{fig:pol_rho2}
\vspace{0.1in}
\end{figure}

The polarization spectra plotted in Figure \ref{fig:pol_rho1} demonstrate the effects of modifying the distribution of gas in the disk on the resulting polarization. To better visualize the evolution of the spectrum as the density changes, the differences in two key polarization quantities are plotted in Figure \ref{fig:pol_rho2}. These two quantities are the polarization level at a particular wavelength, in this case the V-band, and the change across the jump in polarization at one of the hydrogen series limits, in this case the Balmer limit. As we will show, comparing the V-band polarization level and the change in polarization at the Balmer limit is useful in tracing the changes to the continuous polarization spectrum as we modify the distribution of gas in the disk.

In varying the density of the disk at the stellar surface, the resulting spectra are easily explained and illustrate an important optical depth effect. As $\rho_0$ increases, the number of electrons available for Thomson scattering increases. As expected, this translates into an increase in the polarization level. While the optical depth through the disk is small, the polarization spectra remain flat, reflecting a wavelength-independent Thomson scattering signature unaltered by hydrogen absorption. As $\rho_0$ increases, it eventually attains a density at which hydrogen absorption of pre- or post-scattered photons imprints the wavelength-dependent opacity signature on the emergent spectrum of polarized light. For a power-law index of 3.5, this occurs roughly when  $\rho_0 = 1.0 \times 10^{-11}$ g cm$^{-3}$. Above this density, the increase in scattering yields greater increases in the polarization level at wavelengths where the absorptive opacity is lowest (longward of the series limits) than at wavelengths where the absorptive opacity is highest (shortward of the series limits). This is clearly exhibited in Figure \ref{fig:pol_rho1} by the appearance of jumps in the polarization spectrum at the Balmer and Paschen limits, and the steep rise of the Balmer jump as shown in Figure \ref{fig:pol_rho2}.

\begin{table}
\begin{center}
\caption{Stellar Parameters}
\label{tab:param1}
\begin{tabular}{lccccc}
 \tableline \tableline
Spectral & Radius & Mass & Luminosity & $T_{\rm eff}$ & $\log(g)$\\
Type &($R_{\sun}$)&($M_{\sun}$)&($L_{\sun}$)&(K)& ($\rm cm\, s^{-2}$)\\
\tableline
B0V & 7.4 & 18 &  4.0 $\times 10^{4}$ & 3.0 $\times 10^{-4}$ & 3.9\\
B2V & 5.3 & 9.1 &  4.8 $\times 10^{3}$ & 2.1 $\times 10^{-4}$ & 3.9\\
B5V & 3.9 & 5.9 &  7.3 $\times 10^{2}$ & 1.5 $\times 10^{-4}$ & 4.0\\
B8V & 3.0 & 3.8 &  1.4 $\times 10^{2}$ & 1.1 $\times 10^{-4}$ & 4.1\\
\tableline
\end{tabular}
\tablecomments{Based on values from \citet{cox00}.}
\end{center}
\end{table}

\subsection{Inclination}

\begin{figure}
\epsscale{1.0}
\plotone{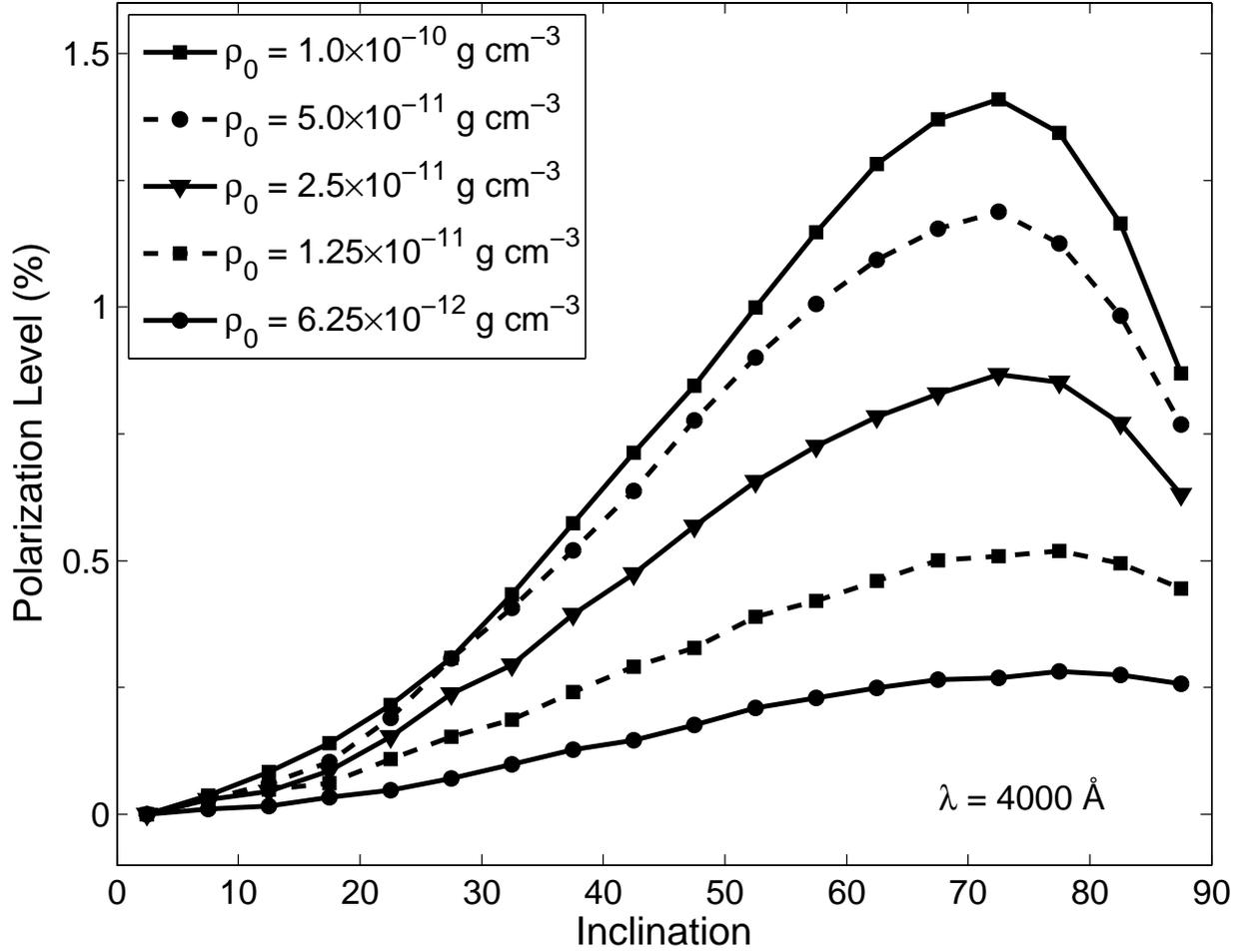}
\caption{Polarization level versus inclination for a B2V star with circumstellar disks with varying $\rho_0$ and $n = 3.5$. For all the values of $\rho_0$ plotted, the polarization level peaks between $i$ = $70^{\circ}$ and $i$ = $80^{\circ}$.}
\label{fig:pol_i}
\vspace{0.1in}
\end{figure}

The inclination of the classical Be star system is critical for the observed polarization. As previously discussed, as the inclination of the system tends to zero, or when the system is viewed pole-on, the fraction of polarized light drops to zero due to the uniform distribution of polarizing planes. One might expect that the polarization would attain its maximum value when the system is viewed edge-on such that the viewing angle coincides with the scattering plane of the disk. At this inclination, we can certainly expect the smallest amount of cancellation in a geometrically thin disk. At high inclinations, however, the scattered light must travel through more gas before it exits the system. A fraction of this polarized light will be absorbed and the overall polarization level will decrease as a result. As Figure \ref{fig:pol_i} demonstrates, the maximum polarization level does not occur at $i$ = $90^{\circ}$, but instead occurs between $i$ = $70^{\circ}$ and $i$ = $80^{\circ}$.

\subsection{Disk Temperature}

\begin{figure*}
\epsscale{1.0}
\plotone{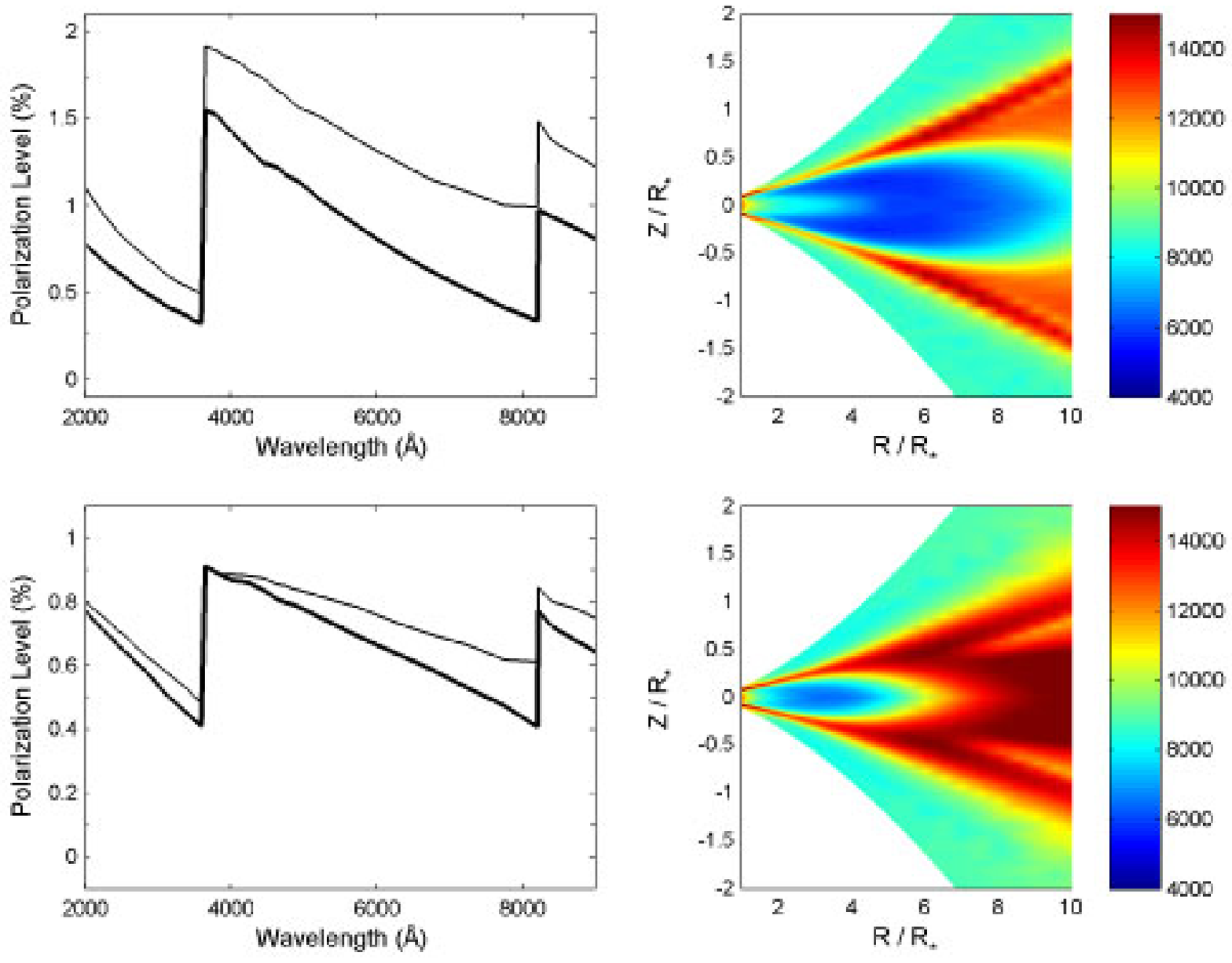}
\caption{Left: Polarization spectra for models using a non-isothermal disk (thin lines) and an isothermal disk (thick lines) viewed at $i$ = $75^{\circ}$. Right: Temperature structure of the non-isothermal disks computed by {\sc bedisk}. The models plotted are for B2V stars surrounded by circumstellar disks with $n = 3.5$ and with $\rho_0 = 1.0 \times 10^{-10}$ g cm$^{-3}$ (top) and $2.5 \times 10^{-11}$ g cm$^{-3}$ (bottom).}
\label{fig:pol_T}
\vspace{0.1in}
\end{figure*}

Recent theoretical studies of classical Be stars using realistic radiative transfers models have emphasized the importance of correctly determining the thermal structure of the disk when modeling observables. Clearly, the disk models produced by \citet{car06a} and \citet{sig07} exhibit temperature maximums and minimums that deviate significantly from an isothermal disk. In order to examine the importance of the temperature structure on the polarization spectrum, we computed the polarization spectra for several models using {\sc bedisk}'s self-consistent thermal structure of the circmustellar disk and compared the output to that from models using an isothermal disk. The isothermal temperatures used were 0.6 $T_{eff}$. A sample of the results of the comparison are illustrated in Figure \ref{fig:pol_T}. The polarization spectra computed using the isothermal models deviate appreciably from the polarization spectra produced using the non-isothermal disk for higher-density models. For optically thinner models, the difference between the polarization levels is much less pronounced. This agrees with the results of a study by \cite{car08}. The computation of the disk temperature solution produces a region of cooler gas in near the equatorial plane of the disk. This affects the level populations in the non-isothermal parts of the disk and an increase in the absorption of scattered photons. Hence, the non-isothermal disks yield smaller polarization levels than the isothermal disks.

\section{Disk Growth and Dissipation}

To illustrate the changes in the polarization during key periods in the evolution of the circumstellar disk, we consider models that approximate the disk growth and dissipation phases. In order to simulate the formation of the disk, we use a series of models with the same values of $n$ and $\rho_0$ and increasing disk size. Similarly, the dissipation of the disk is simulated using a series of models with an inner hole of increasing width. While these series of models represent idealized approximations of actual disk growth and dissipation, the results trace important changes to the observables and demonstrate the possibilities for time-dependent polarimetric measurements as a diagnostic of changing physical conditions in the circumstellar gas.

\subsection{Growth}

\begin{figure}
\epsscale{1.0}
\plotone{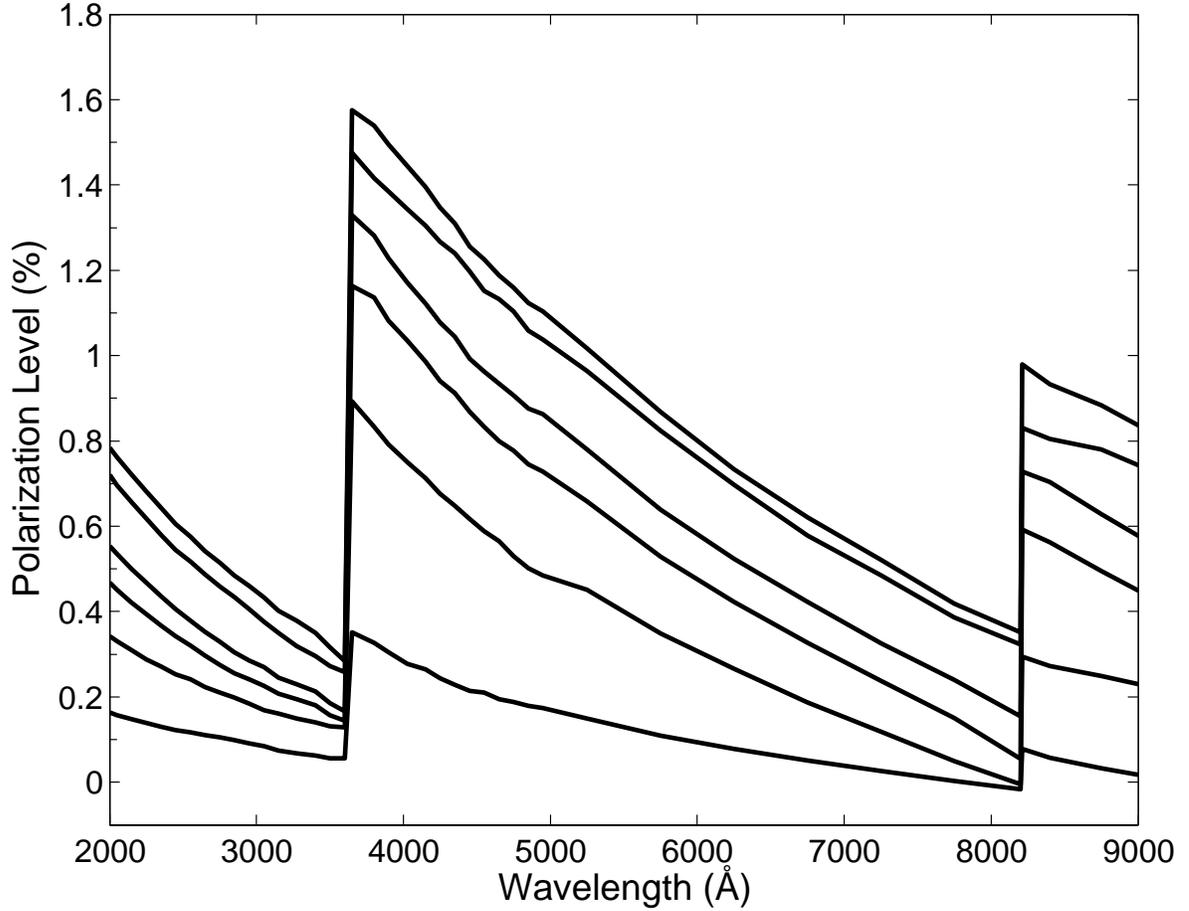}
\caption{Polarization spectra for models with circumstellar disks of increasing outer radius. The models correspond to a disk with $n = 3.5$ and with $\rho_0 = 1.0 \times 10^{-10}$ g cm$^{-3}$ and outer radius, from bottom to top: 2.0 $R_*$, 3.0 $R_*$, 4.0 $R_*$, 5.0 $R_*$, 10.0 $R_*$ and 100.0 $R_*$. The modelled star is a B2V star viewed at an inclination of $i = 75^{\circ}$.}
\label{fig:pol_g1}
\vspace{0.1in}
\end{figure}

\begin{figure}
\epsscale{1.0}
\plotone{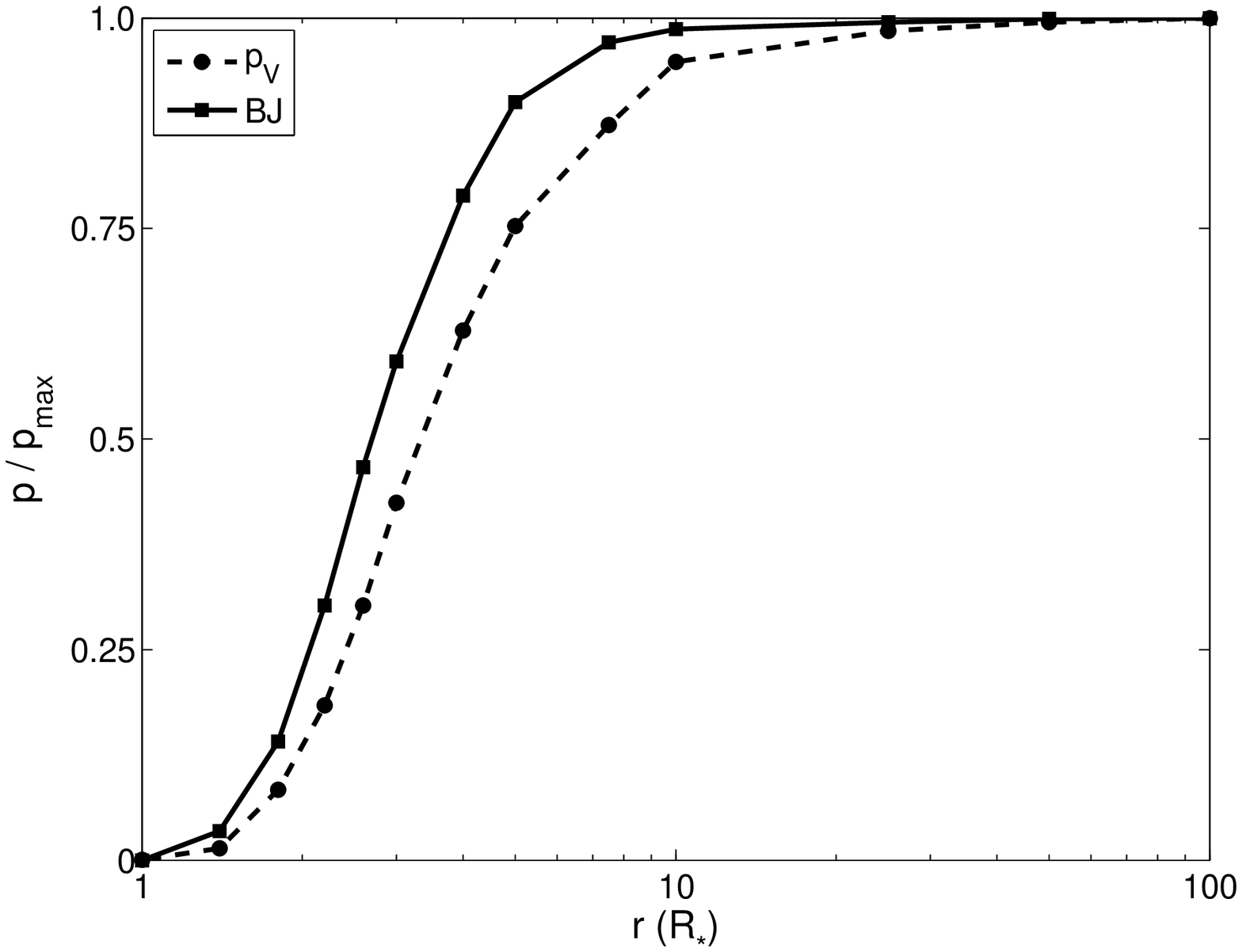}
\caption{Comparison of the growth of polarimetric features for models with circumstellar disks of increasing outer radius. The models correspond to a disk with $n = 3.5$ and with $\rho_0 = 1.0 \times 10^{-10}$ g cm$^{-3}$ and outer radius ranging from 1.0 $R_*$ to 100.0 $R_*$. The modeled star is a B2V star viewed at an inclination of $i = 75^{\circ}$. The differences between the two quantities in the models reflect that gas ejected from the central star is forming an equatorial disk.}
\label{fig:pol_g2}
\vspace{0.1in}
\end{figure}

We produced a series of models that approximates disk growth by incrementally increasing the outer radius of the circumstellar disk from the stellar surface up to 100 $R_*$. With each adjustment to the size of the disk, we recompute the thermal structure of the disk using the radiative transfer code. Figure \ref{fig:pol_g1} shows the gradual change in the polarization spectra as the size of the disk increases. Figure \ref{fig:pol_g2} plots the evolution of the two previously-introduced polarization quantities, the V-band polarization level and the change in polarization at the Balmer jump. 

The formation of the disk is reflected in the polarization spectra with the steep rise of the Balmer jump and the comparatively more gradual rise of the polarization level. As mass is ejected from the stellar surface, the density and optical depth in the innermost region of the disk rapidly increases to the point at which the wavelength-dependent imprint of the absorptive opacity appears in the polarization spectrum. As the disk increases in mass and extent, the number of scatterers increases and the solid angle subtended by the disk grows; hence, the polarization level rises, albeit more slowly than the Balmer jump. The differences in the changes of these two quantities are reflective of a system in which equatorial mass loss from the central star builds a geometrically thin, gaseous disk. 

\citet{car11,car12} has emphasized the importance of careful consideration of the location of the disk where observables originate when analyzing observations. Knowledge of the formation region for different observables is essential in developing diagnostics that trace the evolution of physical changes in the disk. Using the disk growth models as a basis for determining the disk formation regions, we corroborate Carciofi's assertion that 95 \% of the maximum V-band polarization level is reached when the disk size reaches 10 stellar radii. We note that the Balmer jump reaches this level at much smaller disk radii. It attains 95 \% of its maximum value when the disk size reaches 6 stellar radii. This difference in the formation regions between the two polarimetric quantities highlights their usefulness as tracers for disk growth and loss.

\subsection{Dissipation}

It is suggested that the dissipation of the circumstellar disks of classical Be stars may begin from the inner disk and proceed outward, leaving an evacuated region close to the star. \citet{wis10} presented spectropolarimetric and H$\alpha$ spectroscopic observations of two Be stars, $\pi$ Aqr and 60 Cyg, with extended temporal coverage that witnessed both disk-loss and disk-renewal phases. During the disk-loss event of 60 Cyg, they noted a lag of the maximum in the H$\alpha$ equivalent width from the maximum V-band polarization.  Similarly, during the disk-loss event of $\pi$ Aqr, they noted a lag in the onset of the minimum in the H$\alpha$ equivalent width strength from the minimum V-band polarization. The authors propose that these observations indicate that the evacuation of the disk proceeded in an "inside-out" manner. 

\begin{figure}
\epsscale{1.0}
\plotone{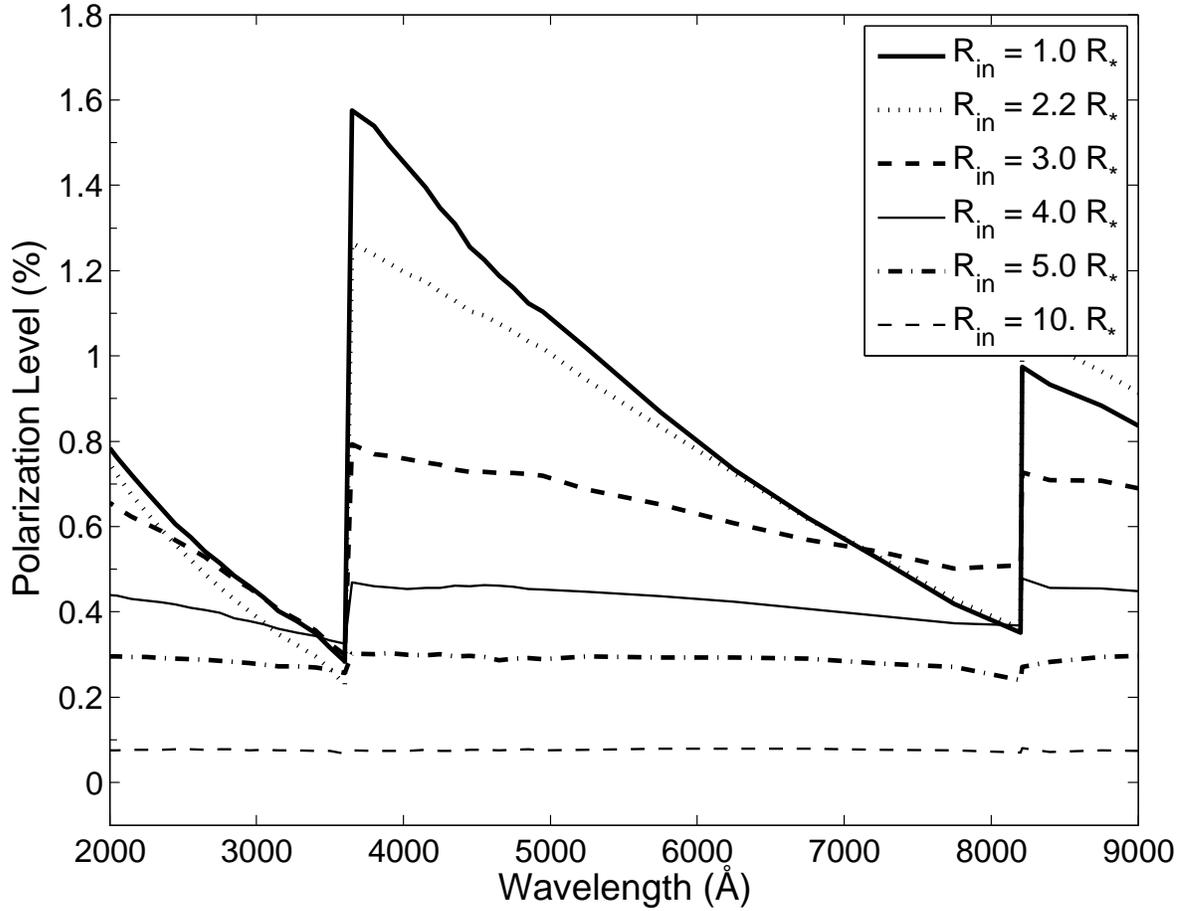}
\caption{Polarization spectra for models with circumstellar disks of increasing inner hole radius. The models correspond to a disk with $n$ and with $\rho_0 = 1.0 \times 10^{-10}$ g cm$^{-3}$. The modeled star is a B2V star viewed at an inclination of $i = 75^{\circ}$.}
\label{fig:pol_d1}
\vspace{0.1in}
\end{figure}

\begin{figure}
\epsscale{1.0}
\plotone{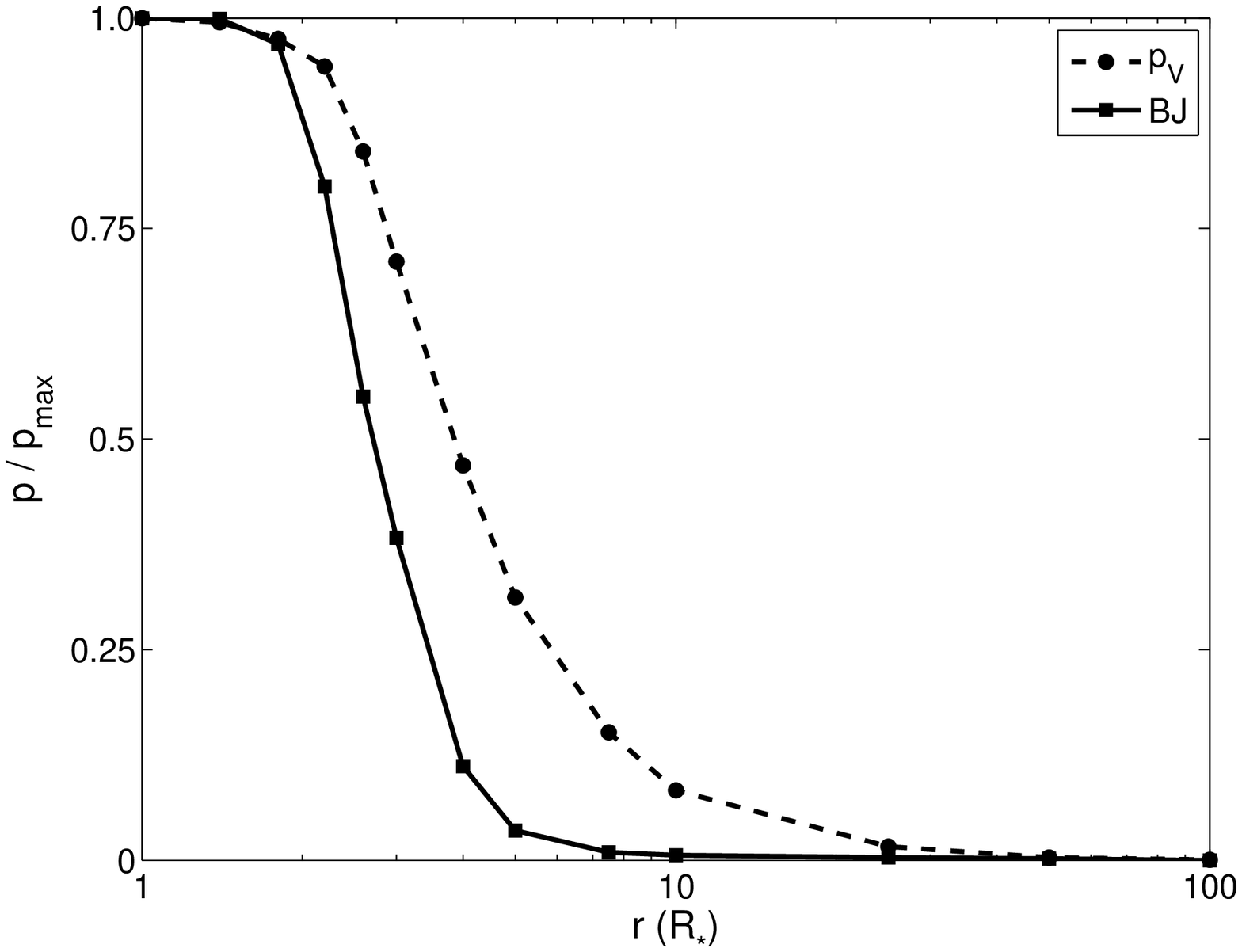}
\caption{Comparison of the change in the polarimetric features for models with circumstellar disks of increasing inner hole radius. The models correspond to a disk with $n = 3.5$ and with $\rho_0 = 1.0 \times 10^{-10}$ g cm$^{-3}$ and inner hole radius ranging from 1.0 $R_*$ to 100.0 $R_*$. The modeled star is a B2V star viewed at an inclination of $i = 75^{\circ}$. The differences between the two quantities in the models reflect that the circumstellar gas is clearing from the inside-out.}
\label{fig:pol_d2}
\vspace{0.1in}
\end{figure}

In order to examine the effect of this process on the polarization level, we produced a series of models with an evacuated inner region. We begin with a disk that extends from the stellar surface up to 100 $R_*$. We introduce an evacuated inner region from the stellar surface up to 1.4 $R_*$ and we incrementally increase the radius of the inner hole with each successive model. With each adjustment to the size of the inner hole, we recompute the thermal structure of the disk using the radiative transfer code. Figure \ref{fig:pol_d1} shows the gradual change in the polarization spectra as the radius of the hole increases. Figure \ref{fig:pol_d2} plots the evolution of the V-band polarization level and the change in polarization at the Balmer jump.

As one would expect, the characteristics that we noted in the growth model are inverted in this series of models. Once mass decretion from the central star terminates, the material in the disk is removed through reaccretion or outward flow. As the dense gas in the innermost region of the disk disappears and the optical depth through the disk diminishes, the height of the Balmer jump declines quickly. Once the disk is cleared up to five stellar radii, the remaining gas is entirely optically thin to hydrogen bound-free absorption and the wavelength-dependent signature of the polarization spectrum vanishes. The V-band polarization gradually declines as the mass of the disk and the number density of electrons decreases. Again, the differences in the changes of these two quantities in this series of models are reflective of the physical change to the circumstellar gas: it is being cleared from the inside-out.

\subsection{BJV Loop}

\begin{figure}
\epsscale{1.0}
\plotone{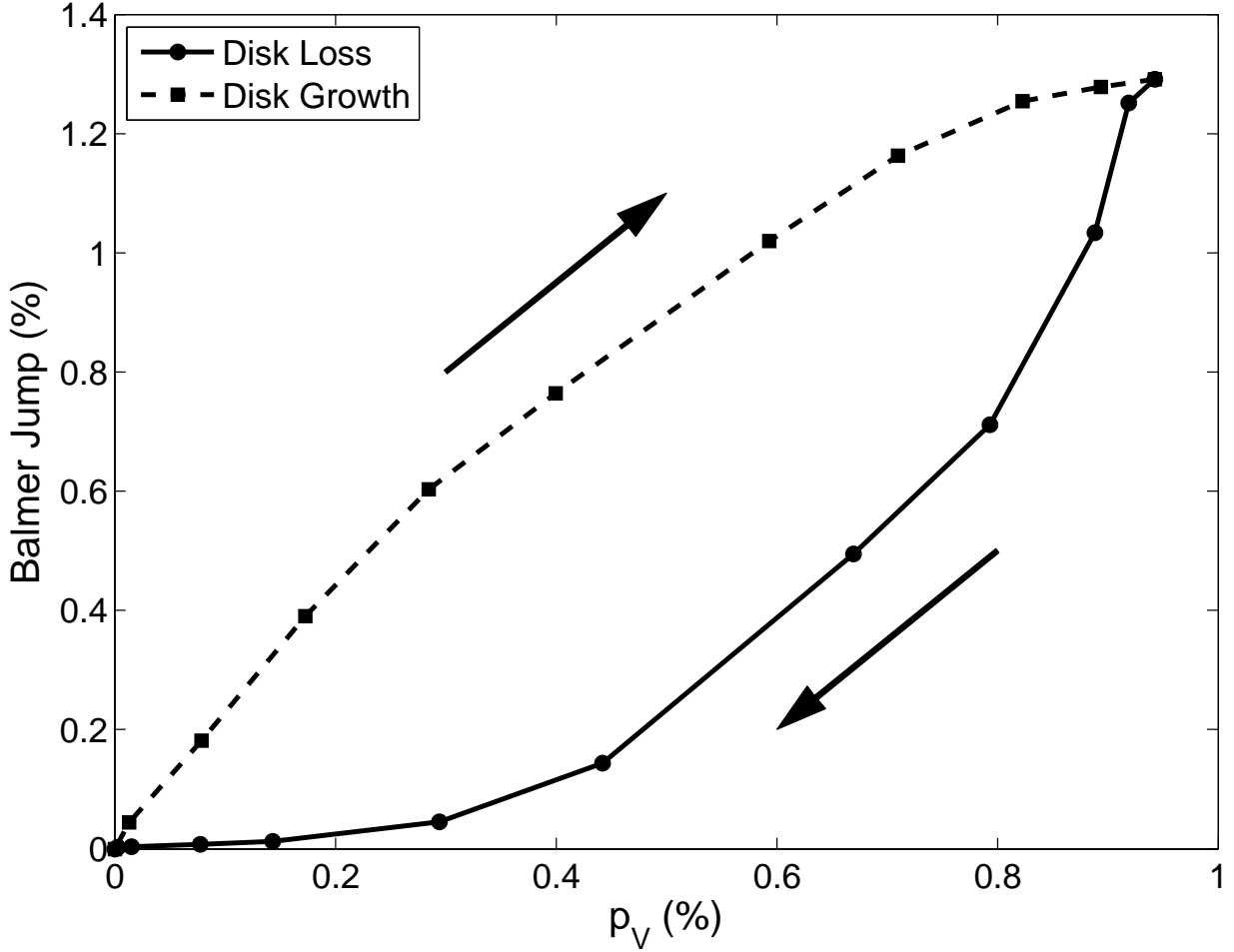}
\caption{BJV loop using the models presented in sections 5.1 and 5.2.  The modelled star is a B2V star viewed at an inclination of $i = 75^{\circ}$ and surrounded by a disk with $n = 3.5$ and with $\rho_0 = 1.0 \times 10^{-10}$ g cm$^{-3}$. The disk dissipation models correspond to an inner hole of increasing radius ranging from 1.0 $R_*$ to 100.0 $R_*$. The disk growth models correspond to an increasing outer radius ranging from 1.0 $R_*$ to 100.0 $R_*$. The evolution of the system, from the moment that mass decretion from the central star begins until the moment where the disk is completely dissipated, traces a clockwise loop in the BJV diagram.}
\label{fig:pol_loop}
\vspace{0.1in}
\end{figure}

\begin{figure*}
\epsscale{1.0}
\plotone{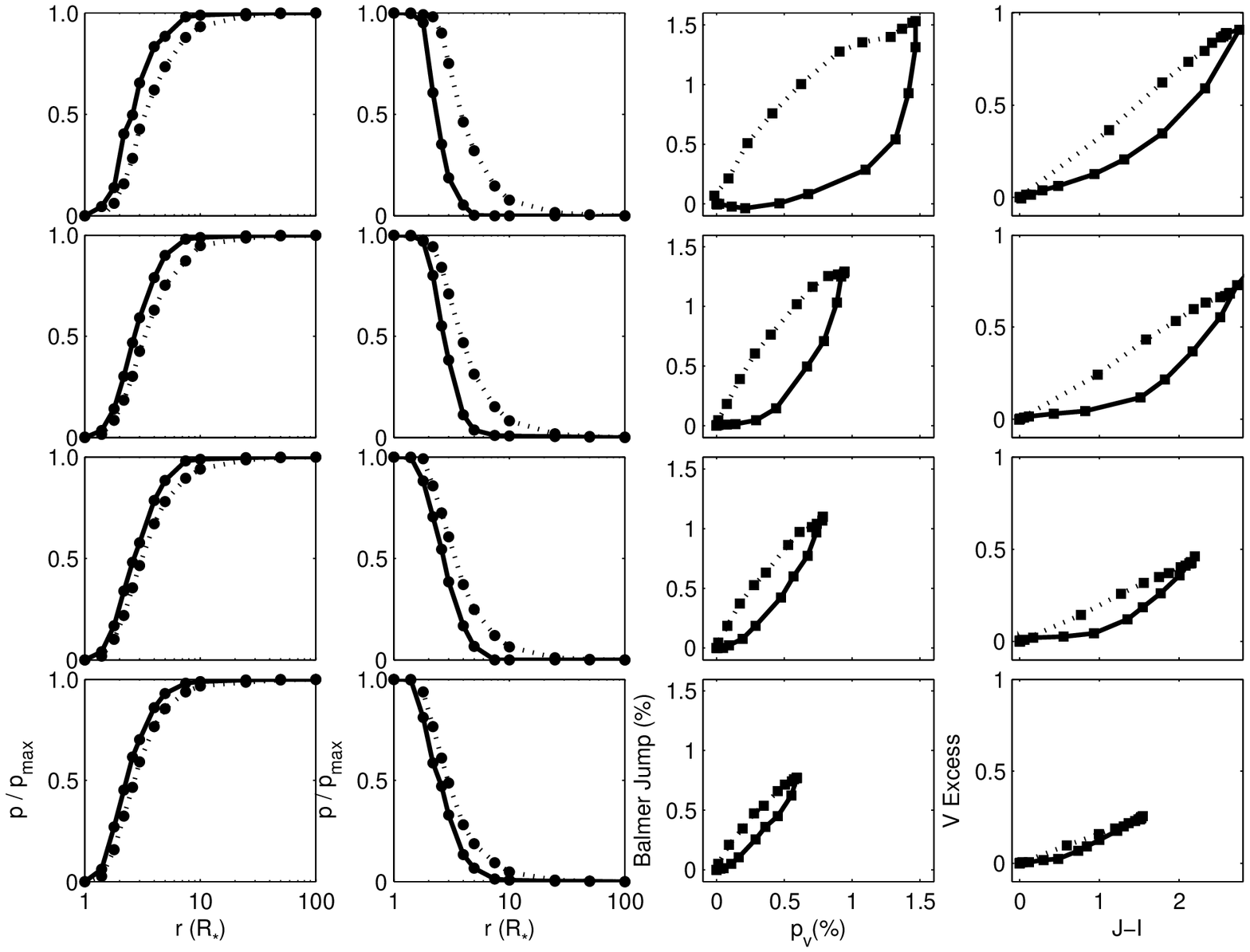}
\caption{Column 1: Balmer jump (solid line) and V-band polarization level (dotted line) during disk growth. Column 2: Balmer jump (solid line) and V-band polarization level (dotted line) during disk dissipation. Column 3: BJV diagram at $i = 75^{\circ}$ showing disk growth (dotted line) and dissipation (solid line). Column 4: Color magnitude diagram at $i = 15^{\circ}$ showing disk growth (dotted line) and dissipation (solid line). From top to bottom: B0V, B2V, B5V and B8V stars. The models correspond to disks with $n = 3.5$ and with $\rho_0 = 1.0 \times 10^{-10}$ g cm$^{-3}$.}
\label{fig:pol_loop2}
\vspace{0.1in}
\end{figure*}

With the results from our disk growth and disk dissipation approximations, we can illustrate the complete cycle of the evolution of the disk using the Balmer jump and the V-band polarization level in what is referred to as a BJV diagram \citep{dra11}. In Figure \ref{fig:pol_loop}, we plot the BJV diagram for the system modelled in the previous two sections. The progression of the system, as the disk grows from the stellar surface out to 100.0 $R_*$ and then dissipates outward from the stellar surface until the disk has completely disappeared, traces a clockwise loop in the BJV diagram. Qualitatively similar loops have been observed in color magnitude diagrams of Be stars \citep{dew06}. These loops are an illustration of the effects of changes in the distribution of gas in the circumstellar disk for observables that arise from different parts of the disk. In Figure \ref{fig:pol_loop2}, we illustrate theoretical loops in color magnitude diagrams and BJV diagrams for four disk models.

Having presented the polarimetric characteristics of models that approximate a Be disk during formation and dissipation, we now examine the physical trends exhibited in BJV diagrams. We have already stated that the electron scattering polarization level, and consequently the level at wavelengths where the absorptive opacity is minimal, is determined mainly by the number of scatterers in the disk. The difference in the V-band polarization level between disks with identical density distributions but surrounding stars with different spectral types can be explained by the factor primarily determining the number of scatterers: the disk temperature. The disks around earlier spectral type stars are hotter and fully ionized. The disks around later spectral types are cooler and less ionized. As such, the disks surrounding earlier spectral type stars are more amenable to electron scattering and produce higher unattenuated polarization levels.

\begin{figure}
\epsscale{1.0}
\plotone{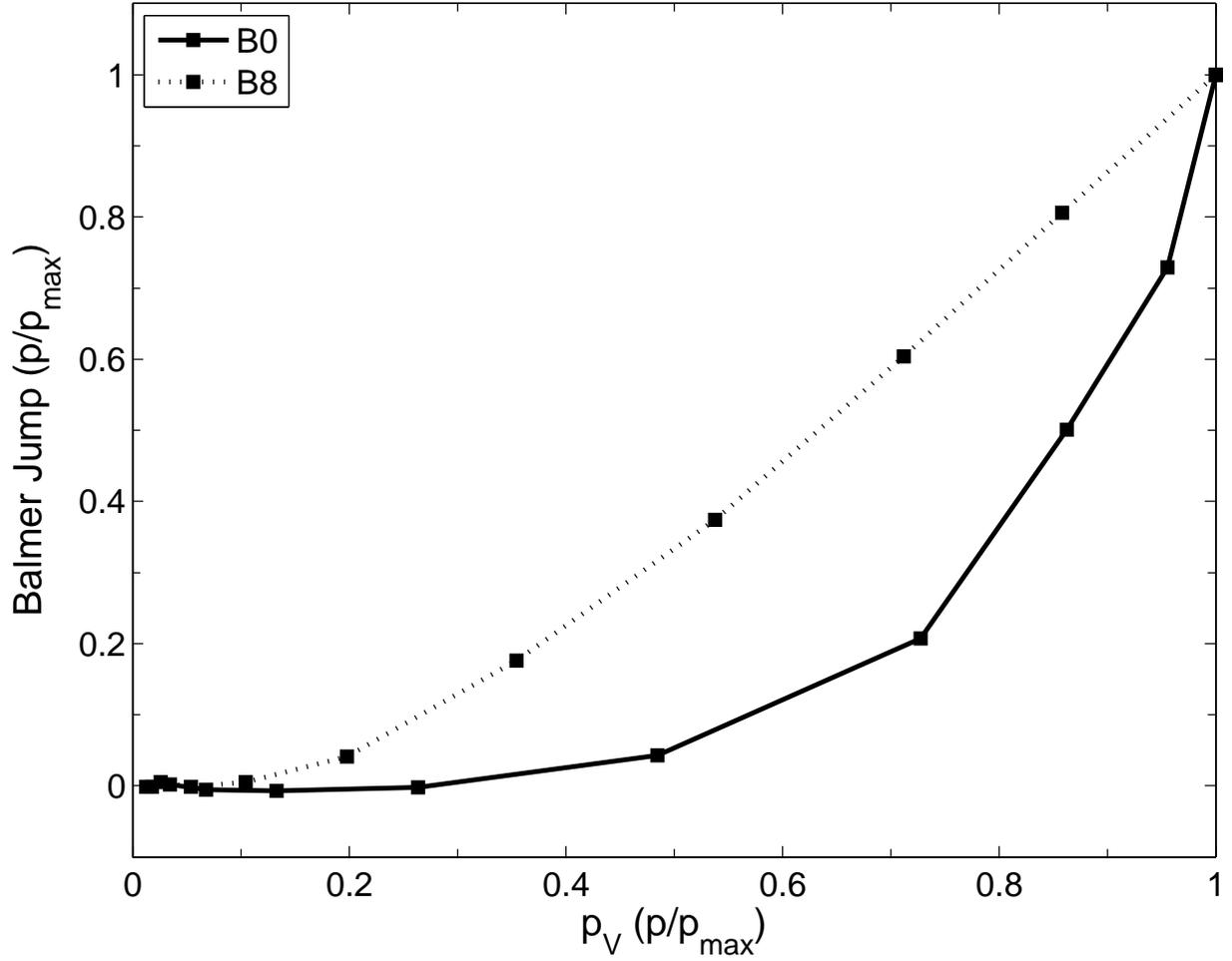}
\caption{BJV curves for disks of increasing mass. The modelled stars are viewed at an inclination of $i = 75^{\circ}$ and surrounded by a disk with $n = 3.5$ and with $\rho_0$ increasing from $3.81 \times 10^{-13}$ g cm$^{-3}$ to $2.0 \times 10^{-10}$ g cm$^{-3}$.}
\label{fig:pol_loop3}
\vspace{0.1in}
\end{figure}

We now consider the polarization levels at wavelengths where the absorptive opacity is greatest. To analyze the depolarizing effect of absorption by neutral hydrogen, we consider a series of hydrogen-only models of varying $\rho_0$. This allows us to examine the effect of changing the amount of disk material without the geometrical effects inherent to the growth and dissipation models. Figure \ref{fig:pol_loop3} illustrates that the BJV curves produced using these models mirror those of the disk dissipation models. That is, for decreasing disk mass, we observe a clockwise descent from maximum to minimum BJV signature, with the curve more pronounced for the disk surrounding the earlier spectral type star.

\begin{figure*}
\epsscale{1.0}
\plotone{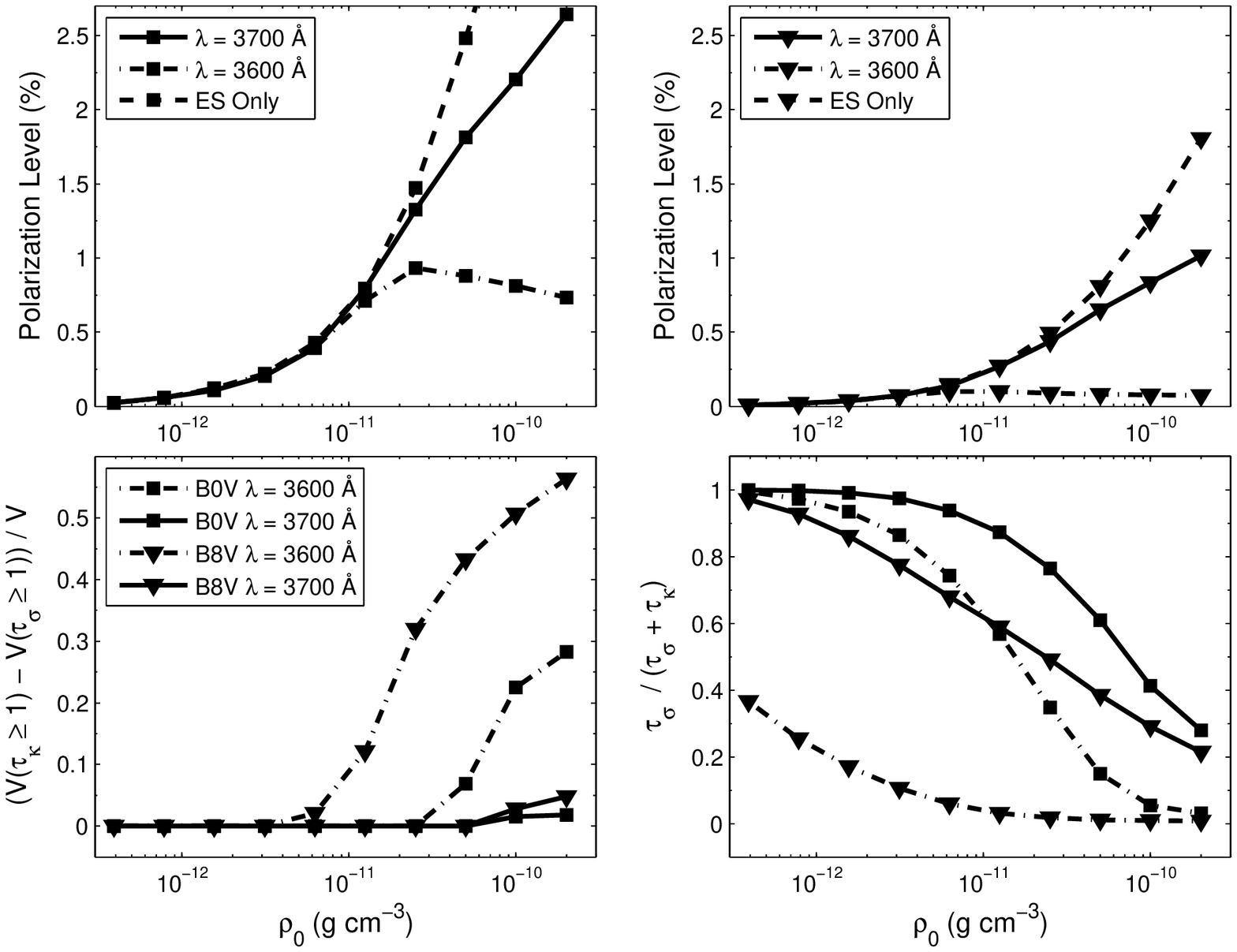}
\caption{Top left: Polarization levels for a B0V star. Top right: Polarization levels for a B8V star. Bottom left: Size of the effective absorbing volume minus the effective scattering volume. Bottom right: Equatorial electron scattering optical depth with respect to the total equatorial optical depth. The modelled stars are viewed at an inclination of $i = 75^{\circ}$ and surrounded by a disk with $n = 3.5$ and with $\rho_0$ increasing from $3.81 \times 10^{-13}$ g cm$^{-3}$ to $2.0 \times 10^{-10}$ g cm$^{-3}$. }
\label{fig:pol_char}
\vspace{0.1in}
\end{figure*}

The top panels in Figure \ref{fig:pol_char} show the the polarization level shortward ($\lambda$ $\approx$ 3600 \AA) and longward ($\lambda$ $\approx$ 3700 \AA) of the Balmer jump for disks surrounding B0V (top-right) and B8V (top-left) stars. As the disk mass increases, the polarization spectrum becomes wavelength-dependent and the levels diverge. This occurs at a lower density for the B8 model, as the volume of the disk where scattering at 3600 \AA~occurs becomes dominated by absorption at lower densities in the cooler disk than in the hotter disk, as shown in the bottom-left panel of Figure \ref{fig:pol_char}. For both stars, the scattering and absorbing regions at 3700 \AA~remain comparable, resulting in changes to the polarization levels that are consistent with the corresponding increase in the number of scatterers. Hence, the jumps in the polarization spectrum are mainly caused by the scattered photons of the disk becoming obscured as they propagate through a larger volume of the disk where the absorptive opacity from neutral hydrogen dominates. 

The general implications of these results with respect to understanding the physical insight in BJV loops are as follows. First, where the slope of the BJV curve is steepest, the absorptive opacity of the disk is changing without much change to the overall amount of polarization from electron scattering. In other words, material in the densest regions of the disk is undergoing significant changes. In terms of what is happening in the disk, this may represent an appreciable mass injection or loss in the inner disk, depending on the direction of the BJV curve. Second, where the slope of the BJV curve is flattest, the unattenuated polarization level is increasing or decreasing without much change in the disk opacity. Hence, the mass of the scattering region is changing, whether from the spread of material from the inner disk to the outer disk (top part of the disk growth curve) or the dissipation of the outer disk once the interior has vanished (lower part of the disk dissipation curve). From this interpretation of the BJV loop, it becomes clearer how this tool may be useful in understanding decretion trends in classical Be stars.

\section{Summary}

The first objective of this project was to develop a code capable of producing synthetic spectra in all four Stokes parameters to supplement a non-LTE radiative transfer computation of the thermal structure of a gaseous circumstellar disk. We report that the computational method that we have developed has been successfully tested and we are using it to analyze and model observations of classical Be stars. In achieving this goal, we have addressed one of the immediate needs identified at the recent workshop on stellar polarimetry held at the University of Wisconsin-Madison \citep[see][]{hof12}, namely to have realistic modeling codes readily available for polarimetric analysis of observations. With time-dependent analysis of the formation and dissipation of their circumstellar disks possible, we encourage the community to engage in polarimetric monitoring of classical Be stars, especially of those objects with known variability.

We have discussed the geometric and physical properties of circumstellar disks that need to be considered when interpreting polarimetric observations of classical Be stars. In particular, we emphasized the use of two quantities, the polarization level in the V-band and the polarization change across the Balmer series limit, as diagnostics of the physical structure of the disk. While the polarization level is a gauge of the number density of scattering electrons in the disk, the Balmer jump indicates the regions where bound-free opacity is greatest. We have shown that these two polarimetric quantities originate from different parts of the disk: the Balmer jump arises from the innermost regions of the disk while the source of the polarization level extends further out in the disk. In particular, we have shown here that the Balmer jump in the polarization spectrum arises from the region of the disk located up to 6 stellar radii from the star. Showing that the excess continuum flux in the K-band arises from a region of this exact radial size, \citet{hau12} stated the importance of this region as being where the highest density variations for periodic decretion are observed. Given our understanding of the dependence of the polarimetric Balmer jump on density, it is clear that this polarimetric feature will be useful for monitoring mass decretion in classical Be stars.

We presented radiative equilibrium consistent models that approximate the formation and dissipation of a thin, decretion disk of gas surrounding a hot, massive star. While the evolution of the disk idealized, these models represent an excellent test bed for predicting and understanding the behaviour of the continuous polarization spectrum during the evolution of the circumstellar disk. We corroborated the theoretical BJV loop presented in \citet{dra11}. In Figure \ref{fig:pol_loop2}, we presented different models which emphasize the difference in the formation region of different observables, a characteristic manifested by loops in BJV diagrams. With this focused study on the polarimetric properties of classical Be stars, we have established a framework for characterizing the BJV loops through knowledge of the physical properties of the disk. These results are particularly important as we expect these loops to reflect time dependent mass decretion in classical Be stars. Thus, observing and modeling these features should play a pivotal role in identifying the mass-loss processes which drive the development of circumstellar disks. 

All of the models in this work assume fixed-rate mass decretion from the surface of the central star and axisymmetry in the distribution of the gas in the disk. As such, these idealized models trace out clear, clockwise loops in the color magnitude and BJV diagrams. Observed color magnitude and BJV diagrams of classical Be stars show more complicated behaviour then presented here.  Modelling and interpreting these observations will require the inclusion of varying mass decretion rates and the consideration of non-axisymmetric distributions of disk material. Furthermore, it has been suggested that in circumstellar environments where material is actively accreting onto the central star, we might observe counter-clockwise loops in colour-magnitude and BJV diagrams. Such loops in classical Be stars may indicate that a significant portion of the disk material reaccretes onto the star during the disk dissipation period. However, \citet{dra11} speculate that counter-clockwise loops in BJV diagrams could be caused by non-constant mass decretion rates or more complicated mass ejection processes. Whatever the case, more sophisticated models of disk growth and dissipation are required for tackling these problems. We are currently working on models which included variable mass-loss rates and non-axisymmetric disk density distributions.

It should be clear that polarimetric observations can aid in identifying the physical properties of circumstellar disks and can be useful in constraining the theoretical models that are crucial to correctly interpreting the observations. We are currently continuing our comprehensive investigation of the polarimetric properties of classical Be stars. While this paper focused on a theoretical consideration of the continuous polarization levels in models of circumstellar disks, our current endeavours also include comparisons with polarimetric observations of classical Be stars. Furthermore, we are currently working to produce synthetic linear spectropolarimetric profiles. This addition to the Monte Carlo scattering code will allow us to further study the geometric structure of the disk on spatial scales that cannot be directly imaged. 

\section{Acknowledgements}
RJH and CEJ would like to acknowledge support from the Canadian Natural Sciences and Engineering Research Council.

\end{document}